\documentclass[twocolumn,showpacs]{revtex4}
\usepackage{epsfig}
\usepackage{amssymb,amssymb}

\begin{document}
\title{DNA electrophoresis in designed channels}
\author{Takahiro Sakaue}
\thanks{\emph{e-mail address:} sakaue@yukawa.kyoto-u.ac.jp}
%
%
\affiliation{Yukawa Institute for Theoretical Physics, Kyoto University, Kyoto 606-8502, Japan}
%


\begin{abstract}
We present a simple description on the electrophoretic dynamics of polyelectrolytes going through designed channels with narrow constrictions of slit geometry.
By analyzing rheological behaviours of the stuck chain, which is coupled to the effect of solvent flow, three critical electric fields (permeation field $E^{(per)} \sim N^{-1}$, deformation field $E^{(def)} \sim N^{-3/5}$ and injection field $E^{(inj)} \simeq N^0$, with $N$ polymerization index) are clarified.
Between $E^{(per)}$ and $E^{(inj)}$, the chain migration is dictated by the driven activation process.
In particular, at $E>E^{(def)}$, the stuck chain at the slit entrance is strongly deformed, which enhances the rate of the permeation.
From these observations, electrophoretic mobility at a given electric field is deduced, which shows non-monotonic dependence on $N$. For long enough chains, mobility increases with $N$, in good agreement with experiments. An abrupt change in the electrophoretic flow at a threshold electric field is formally regarded as a nonequilibrium phase transition.
\end{abstract}

\pacs{83.50.-v, 36.20.Ey, 87.14.Gg}

\maketitle

\section{Introduction}
\label{intro}
The fractionation of DNA molecules depending on their length is a required step in many sectors of biological sciences.
A standard technique for this purpose is a gel electrophoresis, in which negatively charged DNA molecules migrate driven by electric field inside the random environment of a gel \cite{gel_ep1,gel_ep2}.
During the driven diffusion, DNAs have to negotiate with the geometrical constraints imposed by the gel, resulting in lower mobility for longer chains.
It is well known, however, that there is an upper limit for the length, beyond which mobility of DNAs become length-independent.

Many efforts have been carried out to improve on this restriction.
Among them, artificially designed devices with micro-patterns have been getting a lot of attention \cite{gel_ep_artificial_gel,gel_ep_hydro_gel,slitET1,slitET2}.
While in gels the environment for the target molecule is random and there is an upper bound for the statistical ``mesh" size set by the mechanical weakness of dilute gels (for example, the maximum mesh size in agarose gels is $\sim 0.3 \mu m$), the geometry of the device with much larger size constriction can be designed in a desired way.
Another advantage is that the microfabricated device is easy to build in integrative separation systems, which would be required in the field of biomolecules analysis in the future.

Cleary, the progress in this area relies on the understanding of dynamics of polymers in such restrictive environments.
While the dynamics without or with weak external bias has been extensively studied\cite{gel_ep_ET1,gel_ep_ET2,simulation_ET1,simulation_ET2}, little is known about the dynamics involving out of equilibrium, large deformation caused by the external field.
Indeed, long polymers are so soft that they could be easily deformed under weak external forces.
Examples are a chain under stretching\cite{Pincus} and a tetherd chain under hydrodynamic flow\cite{Brochard}, etc..
The same also applies for the polymer dynamics in restrictive environments, the consequences of which seem not to have been explored well.

\begin{figure}[ht]
\begin{center}
\includegraphics[width=6cm]{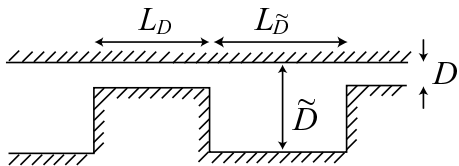}
\caption{A schematic picture of the devise. It consists of a series of wide and narrow slits, whose depth (length) are $\tilde{D}$ ($L_{\tilde{D}}$), $D$ ($L_D$), respectively.}
\label{device}
\end{center}
\end{figure}

In this article, we investigate the electrophoretic dynamics of a polyelectrolyte chain in an artificial channel, where two domains of slit constituting of different depths are alternately arranged (Fig. \ref{device}), with dc electric field.
The choice of this particular geometry is motivated by recent experimental studies, which suggest the potentiality for the DNA separation over the wide range of chain length \cite{slitET1,slitET2}.
A remarkable feature in this system is that longer DNAs migrate faster in contrast to usual situations in gel electrophoresis, which triggered further experimental and several computer simulation studies\cite{slitSM1,slitSM2,slitSM3,slitSM4}.
Here we present a simple scaling argument on the electrophoretic dynamics, which takes the chain rheological behaviours and hydrodynamics interactions into account simultaneously, in the hope of providing lucid descriptions.
An abrupt change of the dynamical state at a certain threshold electric field, as was observed in experiments, is interpreted as a {\it nonequilibrium phase transition} by taking an appropriate limit.
In addition, we briefly discuss the dynamical diagram observed in the gel electrophoresis, in particular the transition from entropic trapping to reptation regimes upon increasing the electric field.

In Sec. \ref{basis}, we briefly remind of the basis of the dynamics of polyelectrolytes under the electric field.
Then, in Sec. \ref{unbias} and \ref{bias}, we analyze the permeation process of polyelectrolytes into the narrow channel.
In Sec. \ref{e_mobility}, we discuss the electrophoretic mobility, in particular, its dependence on the chain length and the electric field.
We predict the four distinct regimes for the electrophoretic dynamics; with the increase in the electric field, one encounters in order, (i) unbiased or weakly biased dynamics, (ii) biased dynamics, (iii) biased dynamics with the significant deformation of the stuck chain, and (iv) free migration without the trapping.
If one relies on the ``local friction" picture (neglect of the hydrodynamics), the regime II disappears, which is discussed in Appendix \ref{case_local_friction}.
To highlight the effect of confinement geometries, we discuss in Appendix \ref{case_cylinder} the electrophoretic dynamics in a cylindrical geometry.
We also address the connection to the flow-injection problem, in which polymers are driven by the solvent flow instead of the electric field (Appendix \ref{connection_flow_injection}).

\section{Dynamics of Polyelectrolytes with Geometrical Traps}
\label{basis}
Consider a polyelectrolyte chain with $N$ segments of length $a$, each of which carrying electric charge of $qe$ ($e$: elementary charge).
For strongly charged polyelectrolytes such as DNA, this charge should be regarded as the renormalized one due to the counterion condensation.
As usual for electrophoresis buffers, the ionic strength of the solution is rather high such that the bulk chain size is given by that of a neutral chain in good solvent, $R \simeq a N^{3/5}$.
It also affects the dynamics through the screening of hydrodynamic coupling on scales larger than Debye length, which leads to the so-called local friction picture, i.e., Rouse dynamics, for the free electrophoresis.
As a consequence, polyelectrolytes in the free solution are treated as free draining coils and the electrophoretic mobility $\mu_{el} = qe/\eta a$ ($\eta$ being the solvent viscosity) is neither dependent on the chain length nor the conformation\cite{gel_ep1}.
However, as was pointed out in ref.{\cite{hydro_screening1}, for polyelectrolytes interacting with an immobile obstacle (more generally, for polyelectrolytes under the simultaneous action of electric fields and non-electric forces), this local friction picture breaks down.
In such situations, one has to deal with the electrohydrodynamics, which is coupled with the deformation behaviours of the chain.
Although this does not allow, in general, a simple analytic treatment, it is possible to linearize the coupled electro-hydrodynamic equations in the case relevant to typical electrophoresis experiments (low Raynolds number situations).
This scheme allows the superposition of electric and non-electric forces, which leads to a hydrodynamic-electric equivalence\cite{gel_ep1,hydro_screening1,hydro_screening2}.
It says that, for example, the deformation of an end-anchored polyelectrolyte under the uniform electric field $E$ is equivalent to that of an end-anchored polymer in a uniform solvent flow at speed $v=\mu_{el}E$.
And the force to immobilize that polyelectrolyte (stall force) $f_{stall} = \gamma \mu_{el} E$, with $\gamma$ being the friction coefficient for the chain in its steady-state conformation.
In the situation of our interest, the polymer may be blocked at the entrance of the narrow slit.
It may be deformed there, but does not show any net motion, while the counterions are still driven without any obstacle, which induces the solvent flow.
Taking these points in mind, in what follows, we consider the dynamics of polyelectrolytes in designed channels by focusing on the elementary step of the channel device, which consists of one narrow slit facing large enough space (Fig. \ref{device}).

\section{Unbiased Dynamics}
\label{unbias}
We first consider the unbiased or weakly biased dynamics, where the effect of electric field is very small or even negligible.
The dynamics of polymers in restrictive environments depends on the relative size ratio between the polymer and the typical size of the confinement, which may be called ``pore".
The situation of our present interest corresponds to the pore size comparable to, or a little smaller than the polymer size in a bulk.
Then, it is known that the free motion (i.e., without external force) of polymers is described by the so-called entropic trapping process \cite{gel_ep_ET1,gel_ep_ET2,simulation_ET1,simulation_ET2}.
The physics involved in this process is essentially the same as size-exclusion chromatography, which states that there is a free energy cost $\Delta F_0(N)$ for the polymer to be squeezed in a narrow space.
It should be noted that even in the situation with non-adsorbing wall in good solvent, the origin of confinement free energy $\Delta F_0(N)$ comes from the segmental interactions or the reduction in the conformational entropy depending on the architecture of the polymer and the geometry of the confinement, which is reflected in the scaling form of $\Delta F_0$ \cite{Sakaue_confinement}.
Since the case of our interest, i.e., a linear polymer confined in a slit, corresponds to the entropic confinement, the terminology of entropic barrier would be appropriate.
In this case, the scaling form of the confinement free energy is \cite{Sakaue_confinement,deGennes} 
\begin{eqnarray}
\frac{\Delta  F_0(N)}{k_BT} \sim \left( \frac{a}{D}\right)^{\frac{5}{3}}N 
\label{F_confine}
\end{eqnarray}
This free energy barrier leads to the exponential slowing down of the chain mobility with the increase in chain length and/or the decrease in the pore size:
\begin{eqnarray}
\mu_0 \sim \exp{\left( -\frac{\Delta F(N)}{k_BT} \right)} 
\end{eqnarray}

\section{Strongly Biased Dynamics}
\label{bias}
If the motion of the chain is biased by the external force, one has to modify the above discussion for the unbiased dynamics.
There are several attempts to this direction including the injection of linear chains \cite{deGennes_review} and branched chains \cite{Sakaue_EPL} into a narrow capillary by hydrodynamic flow, biased electrophoretic mobility of DNA using a slit channel \cite{slitET1,slitET2,slitSM1,slitSM2,slitSM3,slitSM4}.
In these studies, a common strategy is that one notices events occurring inside the channel only, which we call ``inside approach".
The validity of the inside approach can be justified in some cases \cite{deGennes_review,Sakaue_EPL}.
However, the events occurring outside of the channel become important depending on the dissipation mechanism and/or the space dimension of the channel (Appendix \ref{case_local_friction}, \ref{case_cylinder}).
Indeed, we will show that the system under consideration falls upon the latter case.
For the sake of convenience, nevertheless, we first present the inside approach.
The modification due to the outside events is necessary at high electric field, which will be done later. 

\begin{figure}[ht]
\begin{center}
\includegraphics[width=5cm]{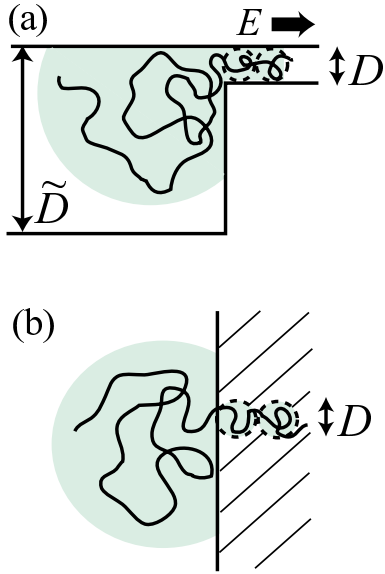}
\caption{A schematic picture of the elementary unit of the channel device ((a)side and (b)top views) and a process of polyelectrolyte permeation by electric field, where the chain section outside the narrow channel is assumed to be unperturbed (``inside approach"). In (b) and FIG. \ref{deform_app}, the shaded part designates the narrow region.}
\label{inside_app}
\end{center}
\end{figure}

\subsection{Inside Approach}
\label{inside_approach}
We now consider the situation, where a large polyelectrolyte chain is facing the entrance of the narrow slit (Fig. \ref{inside_app}).
Then the process of chain permeation may be induced by the electric field.
In the early stage of the permeation, the insertion of a single blob of size $D$ is realized easily.
The corresponding number of monomers is denoted as $g_D$: $D=ag_D^{3/5}$.

But for the chain to be further permeated, a sufficient electrostatic force is required to overcome the osmotic force due to the confinement.
If the moderate range of the electric field is applied, this permeation process is formulated as a barrier crossing problem \cite{slitET1}.
Although there is a two-dimensional freedom for the arrangement of blobs in a slit, the primarily injected blob will proceed straight toward the direction of the electric field, which indeed corresponds to the most energy-saving strategy.
In other words, the nucleated hernia would be more or less straight.
This means that the essential part of the permeation process is described as the one-dimensional problem (Fig. \ref{inside_app}).
This notion is consistent with the video images of ref. \cite{slitET1}.

Let us denote the length of a sequence of blobs already permeated as $y$.
The free energy in the corresponding state is
\begin{eqnarray}
F_{in}(y) = \int_0^y \left( \Pi D^2 -qeE_{in} \frac{y}{a} \right) dy
\label{FreeEnergy_inside}
\end{eqnarray}
The first term in the integrand comes from the confinement effect, where $\Pi \simeq k_BT/(D^3)$ is the osmotic pressure.
The second term represents the force imposed by the electric field inside the slit $E_{in}$.
Equation (\ref{FreeEnergy_inside}) indicates the presence of a free energy barrier at $y=y^*_{in}$:
\begin{eqnarray}
\frac{\Delta F_{in}}{k_BT} \simeq \frac{k_BT}{qeE_{in}} \left(\frac{a}{D^2}\right)
\label{dF_inside}
\end{eqnarray}
and
\begin{eqnarray}
y^*_{in} \simeq \frac{k_BT}{qeE_{in}}\left(\frac{a}{D}\right)
\label{y*_inside}
\end{eqnarray}
The main result of the inside approach with the external force is that the free energy barrier does not depend on the chain length $N$, which is in a remarkable contrast to eq. (\ref{F_confine}).

\subsection{Modification at High Electric Field: Single Chain Rheology}
\label{SingleChainRheology}
The main assumption of the inside approach is the neglect of the chain portion still outside of the narrow channel.
However, this assumption is not always valid in the case of our problem, since as we shall show below, the chain facing the entropic barrier, i.e., the chain stuck at the entrance of a narrow region, is deformed to a notable degree under the influence of the rather high external field.
Then, one has to deal with the single chain rheology; how the chain is deformed under the influence of the external force.
This case with high external force is indeed interesting in the context of the application, which requires the shorter separation time.

Let us first clarify when the deformation of the stuck chain becomes evident.
In experiments, the dielectric constant of the channel is usually much lower than that of the water solution.
Then, the electric flux lines inside the channel interior scarcely leak out.
From the conservation of electric flux and the geometrical symmetry of the channel, the electric field at distance $r$ from the entrance is evaluated as $E(r) \simeq E_{in}D/r$.
In the light of a hydrodynamic-electric equivalence, we calculate the stall force for the stuck chain.
Under the action of the electric field, the chain is pressed against the wall, and its shape is distorted to non-spherical form. 
If the electric field is small enough, it would be plausible to assume that the average chain shape is non-spherical (reflecting the symmetry of the electric field), but the overall size is still given by $R \simeq a N^{3/5}$ (unpertubed size).
Assuming that the stuck chain is not essentially unperturbed (overall chain size is given by $R = a N^{3/5}$), the electrostatic force acting on the whole chain is evaluated as
\begin{eqnarray}
f \simeq \gamma_0 \mu_{el} \bar{E} = \frac{qeE_{in}D}{a}
\label{f_stuck_chain}
\end{eqnarray}
where $\gamma_0 \simeq \eta R$ is the friction coefficient for an unperturbed chain, and $\bar{E} \simeq \int_0^R E(r) r/R^2 dr$ is the average electric field acting on the stuck chain.
If this force is smaller than $k_BT/R$, the assumption of unperturbed chain conformation is valid.
In the opposite case, however, the chain is significantly deformed due to the convergent electric field.
Equating $k_BT/R$ with eq. (\ref{f_stuck_chain}), we have the chain length dependent threshold electric field (in a narrow slit): 
\begin{eqnarray}
E^{(def)} \simeq  \frac{k_BT}{qeD} N^{-\frac{3}{5}}
\label{E_threshold}
\end{eqnarray}
Equation (\ref{E_threshold}) indicates that the longer chain is more easily deformed.

\begin{figure}[ht]
\begin{center}
\includegraphics[width=4cm]{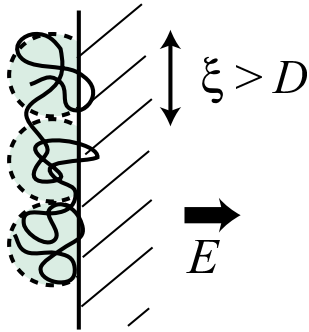}
\caption{A stuck chain at the entrance of the narrow channel under strong electric field ($E^{(def)}<E<E^{(inj)}$). A chain is strongly deformed by the convergent field, and its conformation is represented as a parallel alignment of blobs of size $\xi$. (top view corresponding to FIG. \ref{inside_app} (b)) }
\label{deform_app}
\end{center}
\end{figure}

At $E_{in} > E^{(def)}$, the chain is intensely pressed against the slit entrance, and the molecular shape becomes uniaxial something like a cigar.
The resulting deformed conformation is described as a parallel alignment of blobs of size $\xi$ along the slit entry (Fig. \ref{deform_app}).
Inside blobs, the force due to electric field is not significant, thus, the blob size $\xi$ is related to the number of monomers inside it $g$ as in the bulk, i.e., $\xi = a g^{3/5}$.
Accordingly, blos size $\xi$ is obtained by the following equation
\begin{eqnarray}
\xi \simeq \frac{k_BT}{f_{\xi}}
\end{eqnarray}
where $f_{\xi}$ is the force acting on each blob;
\begin{eqnarray}
f_{\xi}\simeq qeE_{in}D/a
\end{eqnarray}
From these, we obtain
\begin{eqnarray}
\xi \simeq \frac{k_BT}{qeE_{in}} \left(\frac{a}{D}\right)
\label{blob_size}
\end{eqnarray}
Therefore, the length of the stuck chain along the slit entrance is
\begin{eqnarray}
R_{\parallel} \simeq aN \left(\frac{qeE_{in}D}{k_BT} \right)^{\frac{2}{3}}
\end{eqnarray}
If the blob size $\xi$ is smaller than the slit size $D$, there is no entropic barrier arising from the difference in the osmotic pressure; the chain will be injected smoothly inside the slit. 
The critical field for the injection is, thus, obtained  from eq. (\ref{blob_size}) with $\xi =D$ as
\begin{eqnarray}
E^{(inj)} \simeq \frac{k_BT}{qe}\left( \frac{a}{D^2} \right) \simeq E^{(def)} \left( \frac{R}{D}\right)
\label{E_cri}
\end{eqnarray}
At electric field above $E^{(inj)}$, the chain trapping at the slit entrance is irrelevant, therefore, the local friction assumption becomes valid for the description of the electrophoretic dynamics of polyelectrolytes. 
Then the electrophoretic mobility is independent on both the chain length and the electric field.
Note that the same phenomenon for the polyelectrolyte without hydrodynamic interactions corresponds to the observation in gel electrophoresis at high electric field (see Appendix \ref{case_local_friction}, \ref{case_cylinder})\cite{gel_ep_ET2}.
Equation (\ref{E_cri}) provides a simple scaling estimate for the critical field of the abrupt transition on flow behaviours.
For quantitative analyses, however, more details regarding the prefactor are required, which shall be discussed below.

We consider cases with the electric field slightly lower than the critical injection field given by eq. (\ref{E_cri}); $E_{in} = \alpha E^{(inj)}$ ($\alpha < 1$).
Then there still remains a small free energy barrier and an activation process of the permeation will be governed by the following free energy, which is a modified version of eq. (\ref{FreeEnergy_inside}) with the effect of outside events.
\begin{eqnarray}
F(y) = \int_0^y \left( \Delta \Pi D^2 -qeE_{in} \frac{y}{a} \right) dy
\label{FreeEnergy}
\end{eqnarray}
where $\Delta \Pi$ is the difference of osmotic pressures inside and outside of the slit and is given by
\begin{eqnarray}
\Delta \Pi \simeq k_BT \left( \frac{1}{D^3}-\frac{1}{\xi^3} \right)
\end{eqnarray}
Free energy eq. (\ref{FreeEnergy}) takes a maximum value at 
\begin{eqnarray}
y^* \simeq D \frac{(1-\alpha^{3})}{\alpha}
\label{critical_y}
\end{eqnarray}
and the corresponding free energy barrier is
\begin{eqnarray}
\frac{\Delta F}{k_BT} \simeq\frac{(1-\alpha^{3})^2}{\alpha} . 
\label{dF}
\end{eqnarray}
Note that this description is valid at $E_{in} > E^{(def)}$.
At the threshold, $E_{in} = E^{(def)}$, the blob size becomes equal to the unperturbed coil $\xi \simeq R_{\parallel} \simeq R$.
At $E_{in} < E^{(def)}$ the inside approach is valid.

An important consequence of the chain deformation for $E_{in} > E^{(def)}$ is that there are $N/g$ blobs stuck at the slit entry.
Every blob has an equal chance to climb the barrier toward a nucleation.
The transition rate for this activation process is, therefore, estimated as
\begin{eqnarray}
\kappa \simeq \tau_1^{-1} \frac{N}{g} 
\label{transition_rate}
\end{eqnarray}
where $\tau_1 \simeq \tilde{\tau_1} D/\xi$ is the mean escape time from the free energy trap, which is initiated from any single blob; $\tilde{\tau_1} \sim \exp{(\Delta F/k_BT)}$ with $\Delta F$ given by eq. (\ref{dF}).
The factor $D/\xi$ in $\tau_1$ reflects the fact that the larger blob has larger contact area with the slit entrance\cite{slitET1}. 
The escape time is calculated by analysing the stochastic process of the barrier crossing.
From eq. (\ref{FreeEnergy}), the Fokker-Plank equation for this process is
\begin{eqnarray}
\frac{\partial P(y,\ t)}{\partial t} = \frac{\partial}{\partial y}\frac{1}{\gamma (y)}\left[ \frac{\partial F(y)}{\partial y} + \frac{\partial}{\partial y}k_BT\right] P(y,\ t)
\label{F_Peq}
\end{eqnarray}
where, $\gamma (y) \simeq \eta y$ is $y$-dependent friction coefficient.
Evaluation for $\tilde{\tau_1}$ is given as a mean first passage time from $y\simeq0$ to $y_0 >> y^*$\cite{Gardiner}.
By assuming the boundaries at $y=0$ and at $y=y_0 >> y^*$ to be reflective and absorbing, respectively, we obtain
\begin{eqnarray}
\tilde{\tau_1} \simeq \int_0^{y_0}dy\ y\ \exp{\left\{\frac{F(y)}{k_BT}\right\}} \int_0^{y} dy' \frac{\exp{\left\{-\frac{F(y')}{k_BT}\right\}}}{y' \mathcal{D}(y')}
\label{tau_1_exact}
\end{eqnarray}
where $\mathcal{D}(y) \simeq k_BT/\gamma (y)$ is the ($y$-dependent) diffusion coefficient of the partly confined chain\cite{Diff_confined_polymer}.
Following the standard procedures (utilizing the fact that the integrand of the first integral in eq. (\ref{tau_1_exact}) is sharply peaked at $y=y^*$, while that of the latter is very small around there), it is approximated as
\begin{eqnarray}
\tilde{\tau_1} \simeq  t_D \alpha^{-\frac{3}{2}} \exp{\left(\frac{\Delta F}{k_BT}\right)}
\label{tau_1}
\end{eqnarray}
where $t_D =\eta D^3/(k_BT)$ is the characteristic time scale corresponding to the length scale of $D$.
Substituting eq. (\ref{dF}), (\ref{tau_1}) into eq. (\ref{transition_rate}), we obtain
\begin{eqnarray}
\label{transition_rate_1}
\kappa = t_D^{-1} w^{(3)}(\alpha,\ D) N\exp{\left( -\frac{\Delta F}{k_BT}\right)} \\ (E^{(def)}<E_{in}<E^{(inj)})\nonumber
\end{eqnarray}
with 
\begin{eqnarray}
w^{(3)}(\alpha,\ D) \simeq \left(\frac{a}{D}\right)^{\frac{5}{3}} \alpha^{\frac{13}{6}}
\end{eqnarray}

At electric field below $E^{(def)}$, the chain deformation becomes irrelevant.
The free energy barrier is, then, obtained through the inside approach (eq. \ref{dF_inside}) and the transition rate is given by $\kappa = \tau_1^{-1} \simeq (\tilde{\tau_1}D/R)^{-1}$;
\begin{eqnarray}
\label{transition_rate_2}
\kappa = t_D^{-1} w^{(2)}(\alpha,\ D) N^{\frac{3}{5}}\exp{\left( -\frac{\Delta F_{in}}{k_BT}\right)}  \\  (E^{(per)}<E_{in}<E^{(def)})\nonumber
\end{eqnarray}
with
\begin{eqnarray}
w^{(2)}(\alpha,\ D) \simeq \left( \frac{a}{D} \right) \alpha^{\frac{3}{2}}
\end{eqnarray}

When the electric field is further decreased, the critical nucleation size of the hernia, $y^*_{in}$ (eq. (\ref{y*_inside})) becomes longer and equal to the length of the completely confined chain $L=Na^{5/3}/D^{2/3}$ at $E=E^{(per)}$
\begin{eqnarray}
E^{(per)} \simeq \frac{k_BT}{qeD}\left(\frac{D}{a}\right)^{\frac{2}{3}}N^{-1} \simeq E^{(inj)} \left( \frac{D}{R}\right)^{\frac{5}{3}}
\label{E_ub}
\end{eqnarray}
At electric field below $E=E^{(per)}$, the free energy barrier is no longer $\Delta F_{in}$, but given by $F_{in}(L)$, which approaches the confinement free energy $\Delta F_0$ (eq. (\ref{F_confine})) in the limit of zero electric field.
The effect of electric field is, then, very small, and the transition rate is given by
\begin{eqnarray}
\kappa = t_D^{-1} w^{(1)}(\alpha,\ D,\ N) N^{\frac{3}{5}}\exp{\left(-\frac{F_{in}(L)}{k_BT} \right)} \qquad (E<E^{(per)})
\label{transition_rate_3}
\end{eqnarray}
with
\begin{eqnarray}
&&w^{(1)}(\alpha,\ D,\ N) \nonumber \\ 
&\simeq& \frac{a}{D}\left\{\frac{L}{D-\alpha L}-\left(\frac{D}{D-\alpha L} \right)^2 + \frac{1}{\alpha} + \left(\frac{D}{\alpha L} \right)^2\right\}^{-1}
\end{eqnarray}
To obtain eq. (\ref{transition_rate_3}), one has to take notice that the location of the free energy maximum is no longer given by $y=y_{in}^*$ (eq. (\ref{y*_inside})), but given by $y=L$ (the length of the completely confined chain).
In contrast to other regimes for higher electric field, the transition rate becomes exponentially small with increase in the chain length (Sec. \ref{unbias}).

Hereafter, we refer to the migration dynamics at $E<E^{(per)}$, $E^{(per)}<E<E^{(def)}$ and $E^{(def)}<E<E^{(inj)}$ as regime I, regime II and regime III, respectively.

\section{Electrophoretic Mobility}
\label{e_mobility}
After having clarified the physics involved in the elementary step of the device, it is now possible to discuss the mobility of the chain going through an artificial channel consisting of a succession of geometrical traps.
The mobility of the chain could be expressed as \cite{slitET1}
\begin{eqnarray}
\frac{\mu}{\mu_{el}} = \frac{t_0}{t_0+\kappa^{-1}}
\label{mobility}
\end{eqnarray}
The time lapse for a chain to travel an elementary unit is $t_0= (\mu_{el} E_{in})^{-1}(L_D+L_{\tilde{D}}\tilde{D}/D)$, which is a controllable quantity from the device design (see Fig. \ref{device}).
Using eq. (\ref{transition_rate_1}), (\ref{transition_rate_2}) and (\ref{transition_rate_3}), the explicit formula for the mobility is obtained;
\begin{eqnarray}
\frac{\mu}{\mu_{el}} \simeq \frac{t_0}{t_0+AN^{-1}}  \qquad \qquad(E^{(def)}<E<E^{(inj)})
\label{mobility_explicit_1}
\end{eqnarray}
\begin{eqnarray}
\frac{\mu}{\mu_{el}} \simeq \frac{t_0}{t_0+BN^{-\frac{3}{5}} }  \qquad \qquad(E^{(per)}<E < E^{(def)})
\label{mobility_explicit_2}
\end{eqnarray}
\begin{eqnarray}
\frac{\mu}{\mu_{el}} \simeq \frac{t_0}{t_0+(t_D/w^{(1)}(N))N^{-\frac{3}{5}} \exp{\left(\frac{F_{in}(a^{5/3}N/D^{2/3})}{k_BT}\right)}} \label{mobility_explicit_3} \\
 \qquad \qquad(E<E^{(per)} \nonumber)
\end{eqnarray}
where, for simplicity, we show the explicit dependence on $N$ only.

\begin{figure}[ht]
\begin{center}
\includegraphics[width=7cm]{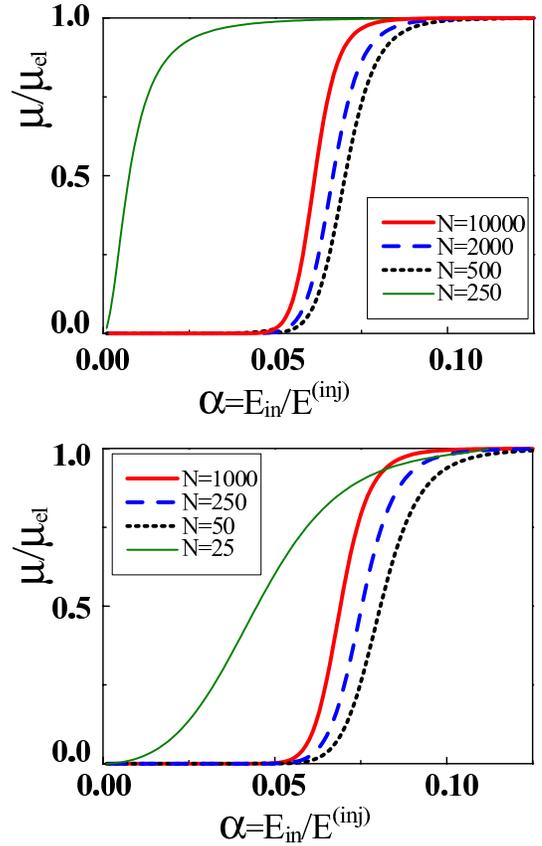}
\caption{Electrophoretic mobility of a polyelectrolyte in a designed channel as a function of electric field with the following parameters: (Top) $D=8 \ a$, $t_D = 5 \times 10^{-9}$ (sec), $t_0 =0.2$ (sec) and $N=200$, $N=500$, $2 \times 10^3$ and $10^4$, (Bottom) $D=2 \ a$, $t_D =10^{-6}$ (sec), $t_0 =5$ (sec) and $N=25$, $N=50$, $250$ and $10^3$. Around the transition region of $\alpha$ shown here, the dynamical behaviours correspond to regime II and regime III for the chain with (Top) $N= 2 \times 10^3$, $500$, (Bottom) $50$ and (Top) $N=10^4$, (Bottom) $N=10^3$, $N=250$, respectively. For the chain with (Top) $N=200$, (Bottom) $N=25$, the transition is realized in the regime I.}
\label{mu_E}
\end{center}
\end{figure}

Figure \ref{mu_E} shows dependences of mobility on applied electric field for various chain lengths with different device parameters (The bottom approximately corresponds to the experimental condition in ref. (\cite{slitET1})).
With the increase on field strength, a crossover is observed from a stuck state to a freely migrating state.
Two trends are perceivable here: 
(i) If the chain is longer than a certain threshold $N^*$, a sharp crossover occurs around the well-defined field strength $E^*$, which is almost length independent. One can recognize, however, a clear trend that the crossover takes place at slightly lower field and becomes sharper with the increase in the chain length.
(ii) For sufficiently short (but still larger than $D$) chains: $N^{(D)}=(D/a)^{5/3} < N < N^*$, on the other hand, the crossover becomes less sharp and is realized at much weaker field.

Therefore, the dependence of the crossover field on the chain length is non-monotonic.
As is clear from our analysis, the second point, i.e., larger crossover field for longer chains, is a consequence of the weakly biased entropic trapping process (regime I), and the observation of the opposite dependence signals a strongly biased dynamics (regimes II and III).
While the effect of chain length is exponential in the regime I, it shows up through a prefactor in regimes II and III.
The threshold length separating above two opposite tendencies is given by 
\begin{eqnarray}
N^*=N^{(per)} (E=E^*) = \left(\frac{D}{a}\right)^{\frac{5}{3}} \alpha^{*-1}
\end{eqnarray}
where $\alpha^{*} =E^*/E^{(inj)}$ and the definition of $N^{(per)}$ is introduced below. 
 
Of great practical interest is how the electrophoretic mobility depends on chain length at a fixed dc electric field.
An appreciable dependence on the chain length is expected if one tunes the electric field in place (slightly smaller than $E^{(inj)}$, see Fig. \ref{mu_E}).
This is exemplified in Fig. \ref{mu} for two different situations.
For comparison, the results obtained from the local friction picture (Appendix \ref{case_local_friction}) are also shown in Fig. \ref{mu_Rouse}.
Note that in these figures, the right graphs approximately correspond to the experimental condition in ref. (\cite{slitET1}).

\begin{figure}[ht]
\begin{center}
\includegraphics[width=8.5cm]{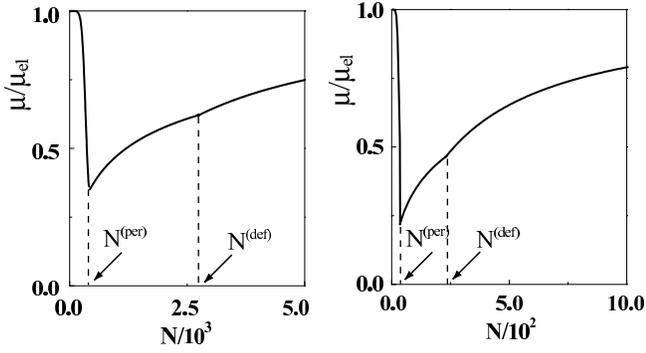}
\caption{Length dependence of the electrophoretic mobility of polyelectrolytes at a given dc electric field with the following parameters: (Left) $D=8 \ a$, $t_D = 5 \times 10^{-9}$ (sec), $t_0 =0.2$ (sec) and $\alpha=0.068$, (Right) $D=2 \ a$, $t_D =10^{-6}$ (sec), $t_0 =5$ (sec) and $\alpha=0.075$. Corresponding characteristic chain lengths are (Left) $N^{(D)} = 32$, $N^{(per)}\simeq 470, N^{(def)} \simeq 2800, N_c \simeq 1200$, (Right) $N^{(D)} \simeq 3$, $N^{(per)}\simeq 40, N^{(def)} \simeq 240, N_c \simeq 270$.}
\label{mu}
\end{center}
\end{figure}

\begin{figure}[ht]
\begin{center}
\includegraphics[width=8.5cm]{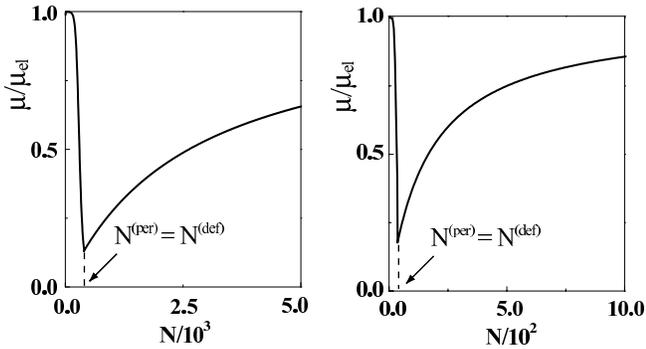}
\caption{Length dependence of the electrophoretic mobility of polyelectrolytes at a given dc electric field calculated according to the local friction model. The parameters are the same as Fig. \ref{mu}. Corresponding characteristic chain lengths are (Left) $N^{(D)} = 32$, $N^{(per)}=N^{(def)} \simeq 470, N_c \simeq 2600$, (Right) $N^{(D)} \simeq 3$, $N^{(per)}=N^{(def)} \simeq 40, N_c \simeq 170$. Around $N=N^{(def)}$, the dynamical regime is shifted from the almost unbiased entropic trap to the biased forced penetration.}
\label{mu_Rouse}
\end{center}
\end{figure}

The electrophoretic behaviours can be classified into the following four regimes.
(0)There is no effect of the geometrical trap for very short chain ($N<N^{(D)} = (D/a)^{5/3}$), then, the mobility is the same as that in the free draining limit.
(I)For short chains ($N^{(D)}<N<N^{(per)}$), the process is described by weakly biased entropic trapping (regime I). The mobility decreases exponentially with the chain length (eq. (\ref{mobility_explicit_3})).
Here, $N^{(per)} \simeq (k_BT/qeD)(D/a)^{2/3}E^{-1}$ from eq. (\ref{E_ub}).
(II)For chains with moderate length ($N^{(per)}<N<N^{(def)}$), the permeation is driven by the electric field, but the conformation of the stuck chain is still in equilibrium (regime II). The mobility increases with the chain length according to eq. (\ref{mobility_explicit_2}).
Here, $N^{(def)}\simeq (k_BT/qeDE)^{5/3}$ from eq. (\ref{E_threshold}).
(III)For longer chains ($N>N^{(def)}$), the chain distortion at the entrance becomes apparent (regime III), which leads to more enhanced mobility for longer chains (eq. (\ref{mobility_explicit_1})).

As a brief summary, a diagram of the electrophoretic dynamics is shown in Fig. \ref{diagram}. 
\begin{figure}[ht]
\begin{center}
\includegraphics[width=6.5cm]{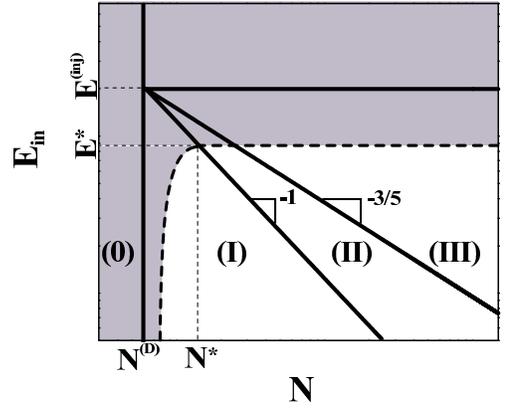}
\caption{A dynamical diagram of the electrophoresis in $(N,\ E_{in})$ plane (Logarithmic scale). Lines separating regimes are: $E^{(per)} \simeq E^{(inj)} (D/R)^{5/3}$ for the border between (I) and (II) and $E^{(def)} \simeq E^{(inj)} (D/R)$ for the border between (II) and (III). The dashed curve between white and gray areas designates a crossover from a stuck to a flowing state, which is obtained from $\kappa t_0 = 1$ (see eq. (\ref{mobility})). The gray area designates the regimes, in which the effect of the geometrical constriction is not apparent because of either too short chain length (left part) or sufficient strong electric field (upper part). At $N >> N^*$, this dashed curve is given by $E^* \sim \left[ \ln{N}\right]^{-1}$ (see eq. (\ref{alpha_transition})) and this crossover exhibits a feature of a nonequilibrium phase transition in the limit of infinite chain length.}
\label{diagram}
\end{center}
\end{figure}
It is worth mentioning a similarity between this diagram and the experimentally observed dynamical diagram of the gel electrophoresis \cite{gel_ep_ET2} (see also Appendix \ref{case_local_friction}, \ref{case_cylinder}).

The chain with $N>N^{(per)}$ is in the driven permeation regime, and the longer chains migrate faster.
This is a consequence of the increased attempt frequency to climb the free energy barrier.
How this attempt frequency depends on the chain length is a matter of the conformational behaviours of the stuck chain, which separates the regime II and III.
We expect that a sufficiently long chain is markedly deformed at the entrance, which enhances the rate of the permeation.

On the other hand, if one neglects the effect of hydrodynamic interactions (i.e., local friction picture), $N^{(per)} = N^{(def)}$ is expected; the regime (II) does not exist.
Therefore, the deformation of the stuck chain is expected to be evident even for chains with moderate length.
The computer simulation studies so far should be compared with this model\cite{slitSM1,slitSM2,slitSM3,slitSM4}.

\section{Discussion}
\subsection{Comparison with Experiments}
First, let us compare our results with the experiment.
In ref. \cite{slitET1}, electrophoresis of DNA was carried out using the narrow slit $D=90$ nm.
Then the sharp crossover from the trapped state to the freely migrating state was found around the electric field $E_{ave}^* \simeq 30$ (V/cm), where $E_{ave}$ is the average electric field over the channel.
From the device parameters used in the experiment, the corresponding electric field in the narrow slit is found to be $E^* \simeq 50$ (V/cm).
The sharpness of the transition (defined as the ratio of the range of the electric field $\Delta E$ between zero and the asymptotic mobility to $E^*$) is also found to be $\Delta E/E^* \simeq 0.6$. 
Putting values corresponding to DNA (segment length $a \simeq 60$ nm and effective charge per segment $q \simeq 85$) into eq. (\ref{E_cri}), we obtain that $E^{(inj)} \simeq 20$ (V/cm), and $\Delta E/E^* \simeq 0.5$ is obtained from Fig. \ref{mu_E} (Bottom).
Thus, we find a good agreement for the sharpness of the transition.
On the other hand, we somewhat underestimate the threshold field, since the experimentally observed value is slightly smaller than $E^{(inj)}$ by a factor of $\alpha^* \sim 0.1$ (see Fig. \ref{mu_E}): Note that the precise value of $\alpha^*$ depends on $t_0$, thus, the device parameters such as $D$, $L_D$ and $L_{\tilde{D}}$.
However, given the overall accuracy of scaling laws and the assumption involved in the simplified model (no leakage of the electric flux at the dielectric boundary, neglect of the effect of electro-osmotic flow), this discrepancy in the threshold field is not surprising.

\subsection{Long Chain Limit: Nonequilibrium Phase Transition}
It may be intriguing to examine dynamic behaviours at the asymptotic limit of the long chain length.
As we shall argue now, a sharp crossover of the electrophoretic flow exhibits features of ``phase transition".
We consider the situation, where the length of the device unit $L_D,\ L_{\tilde{D}}$ is large enough compared to the chain length. (Although this is, of course, not the case for very long chains, one can take the limit, in which the device size is made arbitrary large in proportion to the chain length.)
In this limit, the threshold field separating a stuck and a flowing state becomes
\begin{eqnarray}
\alpha^{*} = \frac{E^*}{E^{(inj)}} \sim \left[\ln{N}\right]^{-1}
\label{alpha_transition}
\end{eqnarray}
and the electrophoretic mobility is expressed as
\begin{eqnarray}
\frac{\mu}{\mu_{el}} \sim \frac{1}{1+const. \ N^{\epsilon}}
\label{mobility_limit}
\end{eqnarray}
where $\epsilon \simeq (\alpha^* -\alpha)/\alpha$ is a reduced electric field.
The crossover becomes shaper and sharper with the increase in $N$ and becomes singular in the limit of infinite chain length as announced.
Therefore, this dynamical state change (from the state with zero to some finite current) caused by the external field is formally regarded as a {\it nonequilibrium phase transition}.

As another possible choice of the limit, one may consider increasing the chain length for a fixed device size.
In such a case, a chain interacts with many elementary device units simultaneously, and some unscripted new features should appear \cite{slitSM4}.

\subsection{Practical Aspects}
A unique property of the present system is that there is a regime where longer chains migrate faster.
This is a consequence of the larger contact area with the slit entrance for larger chains\cite{slitET1}.
For long enough chains ($N^{(def)}<N$), or under the strong electric field ($E^{(def)}<E$), this effect is further enhanced by the electric field induced chain distortion at the geometrical trap, which generates more blobs for longer chains as steady state conformations (regime III).
An increase in blob numbers increases the frequency for the trial to climb the free energy barrier as is clearly seen in eq. (\ref{transition_rate_1}).
If one neglects the effect caused by the solvent flow, the regime II disappears.
Then, the chain stuck at the slit entrance and driven by the electric field is inevitably distorted.
This explains the reason why the stuck chain deformation was observed in Monte Carlo simulations\cite{slitSM1}, while it is not apparent in real experiments using moderate length of DNA\cite{slitET1}.

For very long chains of $N >> N_c = A/t_0$ (or $N_c =  (B/t_0)^{5/3}$ if $(B/t_0)^{5/3} < N^{(def)}$), however, the chain trapping becomes irrelevant ($t_0>>\kappa^{-1}$), and the mobility approaches to the free draining limit.
This bounds the range for efficient separation in a single run of electrophoresis at a given set of parameters.
However, we now know the trend on how this appropriate length range depends on controllable parameters such as size of channels ($D$, $L_D$, $L_{\tilde D}$) and electric field.
The same also applies for the migrating speed of chains, which is another important factor for the practical purpose.
In addition, such microfabricated devices could be easily assembled into practical integrative systems.
Therefore, by constructing multiple channels with designed different conditions, which operate in succession or in parallel, it may be possible to achieve an efficient separation over a wide range of chain length.
This is, in fact, of advantage to use designed channels for separations.

\section{Final Remarks}
Our present analysis is based on the steady state rheological behaviours of the chain.
This approach is motivated by the microscopic images of the experiment\cite{slitET1,slitET2}, in which the transient behaviours (the relaxation after reentering the thick region, the stretching under high electric field gradient, etc.) seem not to be important very well.
However, if the period of the device is made shorter, successive injection and ejection events start to correlate, and the transient dynamics of the chain becomes relevant\cite{slitSM1,slitSM3,slitSM4}.
Also, recent experimental and simulation studies have observed the transition between two distinct states with different migration speeds using relatively deep channels, in which DNAs may be trapped deep in the well and can not escape from it\cite{slitSM3,slitSM4}.
Such features are not studied here, but should be incorporated in the future analysis.

In summary, we have shown that the deformation of the stuck chain and the effect of the flow are important ingredients for the DNA electrophoresis in the designed channel under consideration. 
Such rheological properties in a single chain level are expected to manifest their importance in many other situations in the dynamics of polymers in restrictive environments.

\acknowledgments
I am grateful to E. Rapha\"el, discussion with whom motivated the present work. I am also indebted to T. Ohta for useful discussion and his critical reading of the manuscript. This research was supported by JSPS Research Fellowships for Young Scientists (No. 01263).

\appendix
\section{Electrophoresis based on the Local Friction Assumption}
\label{case_local_friction}
In this appendix, we briefly discuss the electrophoretic dynamics based on the local friction picture, where the role of the solvent flow is neglected.
Although this does not correspond to the real experiments, it provides useful information to analyze the result from computer simulations performed so far.
In addition, this argument may be relevant to the gel electrophoresis, where the hydrodynamics interactions are screened at the scale of the pore size (see also Appendix \ref{case_cylinder}).

Then, the rheology of the stuck chain is modified as follows.
Instead of eq. (\ref{f_stuck_chain}), the electric force acting on the stuck chain is
\begin{eqnarray}
f_{(R)} \simeq \frac{qeE_{in}DN}{R} = \frac{qeE_{in}DR^{\frac{2}{3}}}{a^{\frac{5}{3}}}
\end{eqnarray}
Here and in the followings, the subscript $(R)$ indicates the local friction (Rouse) dynamics.
One finds for the deformation field
\begin{eqnarray}
E_{(R)}^{(def)} \simeq  \frac{k_BT}{qeD} N^{-1}
\label{E_R^def}
\end{eqnarray}
At electric field above $E_{(R)}^{(def)}$, the stuck chain is deformed, and its conformation is a parallel alignment of blob of size $\xi_{(R)}$ (cf. Fig. \ref{deform_app}).
\begin{eqnarray}
\xi_{(R)} \simeq a \left(\frac{k_BT}{qeE_{in}D} \right)^{\frac{3}{5}}
\end{eqnarray}
The length of the stuck chain along the entrance if
\begin{eqnarray}
R_{(R) \parallel} \simeq aN \left(\frac{qeE_{in}D}{k_BT} \right)^{\frac{2}{5}}
\end{eqnarray}
The condition for the smooth chain injection is $D\ge \xi_{(R)}$, which leads to the critical injection filed;
\begin{eqnarray}
E_{(R)}^{(inj)} \simeq \frac{k_BT}{qe}\left( \frac{a^{\frac{5}{3}}}{D^{\frac{8}{3}}} \right) \simeq E_{(R)}^{(def)} \left(\frac{R}{D}\right)^{5/3}
\label{E_inj_R}
\end{eqnarray}
At $E_{in}>E_{(R)}^{(inj)}$, the trapping at the slit entrance is irrelevant, which results in the length and also electric field independent electrophoretic mobility.
This corresponds to the observation in gel electrophoresis that the entropic trapping effect disappears at electric field higher than a certain threshold value\cite{gel_ep_ET2}.
(Note that the topological restriction imposed by gels with rather large pore size may be better modelled by the cylindrical pore geometry, for which we find the same scaling form of the injection field, see Appendix \ref{case_cylinder}.)

The activation process at $\alpha (=E_{in}/E_{(R)}^{(inj)})<1$ can be analyzed similarly.
Now the free energy (eq. (\ref{FreeEnergy})) for the permeation process is modified as
\begin{eqnarray}
F_{(R)}(y) = \int_0^y \left( \Delta \Pi_{(R)} D^2 -qeE_{in}g_D \frac{y}{D} \right) dy
\end{eqnarray}
with
\begin{eqnarray}
\Delta \Pi_{(R)} \simeq k_BT \left( \frac{1}{D^3}-\frac{1}{\xi_{(R)}^3} \right)
\end{eqnarray}
This produces a free energy barrier
\begin{eqnarray}
\frac{\Delta F_{(R)}}{k_BT} \simeq\frac{(1-\alpha^{9/5})^2}{\alpha}
\label{dF_R}
\end{eqnarray}
and its location
\begin{eqnarray}
y_{(R)}^* \simeq D \frac{(1-\alpha^{\frac{9}{5}})}{\alpha}
\end{eqnarray}
By taking into account that the diffusion coefficient of the partly confined chain is now given by $\mathcal{D}_{(R)}(y) \simeq (k_BT/\eta)(a/D)^{2/3}y^{-1}$, the transition rate is calculated as
\begin{eqnarray}
\kappa_{(R)} = t_D^{-1} w_{(R)}^{(3)}(\alpha,\ D) N\exp{\left( -\frac{\Delta F_{(R)}}{k_BT}\right)} \\ E_{(R)}^{(def)}<E_{in}<E_{(R)}^{(inj)} \nonumber
\end{eqnarray}
with
\begin{eqnarray}
w_{(R)}^{(3)}(\alpha,\ D) \simeq \left(\frac{a}{D}\right)^{\frac{7}{3}} \alpha^{\frac{19}{10}}
\end{eqnarray}
Note that this description is valid at $E_{in} > E_{(R)}^{(def)}$ (regime III).
At the threshold, $E_{in} = E_{(R)}^{(def)}$, or equivalently $\alpha \simeq (D/a)^{5/3}N^{-1}$, we have $y_{(R)}^* \simeq (a^{5/3}/D^{2/3}) N =L$, which is the length of the completely confined chain.
Therefore, at this point, we have a crossover to the almost unbiased entropic barrier regime (regime I) and the regime II does not exist ($E_{(R)}^{(per)} \simeq E_{(R)}^{(def)})$.
The transition rate in regime I is calculated as
\begin{eqnarray}
\kappa = t_D^{-1} w_{(R)}^{(1)}(\alpha,\ D,\ N) N^{3/5} \exp{\left( -\frac{F_{in(R)}(L)}{k_BT}\right)} \\
 \ E_{in}<E_{(R)}^{(def)} \nonumber
\end{eqnarray}
where $F_{in(R)}$ describes the free energy of the permeation process based on the inside approach, and
\begin{eqnarray}
&&w_{(R)}^{(1)}(\alpha,\ D,\ N) \nonumber \\ 
&\simeq& \left(\frac{a}{D}\right)^{5/3}\left\{\frac{L}{D-\alpha L}-\left(\frac{D}{D-\alpha L} \right)^2 + \frac{1}{\alpha} + \left(\frac{D}{\alpha L} \right)^2\right\}^{-1}
\end{eqnarray}
The dependence of the electrophoretic mobility on the chain length is represented in Fig. \ref{mu_Rouse}.

\section{Channels with Cylindrical Pores}
\label{case_cylinder}
If the channel is composed of the alternating wide and narrow cylindrical pores, instead of deep and shallow slits, the electrophoretic dynamics of the chain is altered.
In this geometry, the convergent field near the entrance of narrow constriction is $E(r) \simeq E_{in}D^2/r^2$.
The stall force for the unperturbed stuck chain is evaluated as
\begin{eqnarray}
f \simeq \gamma_0 \mu_{el} \bar{E} = \frac{qeE_{in}D^2}{aR}
\label{f_stuck_chain_cylinder}
\end{eqnarray}
where $\gamma_0 \simeq \eta R$ is the friction coefficient for an unperturbed chain, and $\bar{E} \simeq \int_0^R E(r) r^2/R^3 dr$ is the average electric field acting on the stuck chain.
By comparing this force with $k_BT/R$, the deformation field is obtained
\begin{eqnarray}
E^{(def)} \simeq  \frac{k_BT a}{qeD^2}
\label{E_threshold_cylinder}
\end{eqnarray}
In contrast to the slit geometry (eq. (\ref{E_threshold})), $E^{(def)}$ does not depend on the chain length.

At the electric field above $E^{(def)}$, the chain is highly compressed toward the narrow pore and the resultant conformation can be viewed as a dense stacking of blobs, whose size shrinks with approaching to the pore entrance i.e., an {\it inhomogeneous semidilute solution} (Fig. \ref{stuck_chain_cylinder}).
\begin{figure}[ht]
\begin{center}
\includegraphics[width=5cm]{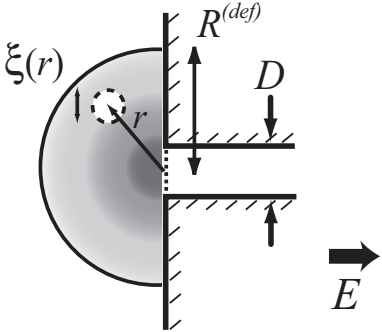}
\caption{A stuck chain at the entrance of the cylindrical pore. The colour strength represents the segment density.}
\label{stuck_chain_cylinder}
\end{center}
\end{figure}
The size of outmost blobs is obtained from
\begin{eqnarray}
\xi(R^{(def)}) \simeq \frac{k_BT}{\gamma_{R^{(def)}} \mu_{el} E(R^{(def)})}
\end{eqnarray}
where $R^{(def)}$ is the overall size of the deformed stuck chain and $\gamma_{R^{(def)}}$ is the friction coefficient of the outmost blob.
\begin{eqnarray}
\xi(R^{(def)}) \simeq \left( \frac{k_BTa}{qeE_{in}D^2}\right)^{\frac{1}{2}}R^{(def)} \simeq \left(\frac{E^{(def)}}{E_{in}}\right)^{\frac{1}{2}}R^{(def)} 
\end{eqnarray}
The force acting on the inner blobs ($r<R^{(def)} $) in a radial direction is the sum of drag forces acting on outer blobs;
\begin{eqnarray}
f(r) \simeq \int_r^{R^{(def)}} \frac{\gamma(r) \mu_{el} E(r)}{\xi(r)} dr \\
\simeq \frac{qeE_{in}D^2}{a}\left(\frac{1}{r}-\frac{1}{R^{(def)}}\right)
\end{eqnarray}
By balancing this force with $k_BT/\xi(r)$, the blob size at $r \ge D$ is expressed as
\begin{eqnarray}
\xi(r) \simeq \frac{k_BT a}{qeE_{in}D^2}\frac{rR^{(def)}}{R^{(def)}-r}
\end{eqnarray}
At $r<D$, the blob does not further shrink any more; $\xi(r) = \xi(D)$ at $r<D$.
The size of the deformed chain $R^{(def)}$ is determined from the segment number conservation;
\begin{eqnarray}
\int_{0}^{R^{(def)}} \frac{g(r)}{\xi(r)^3} d^3r \simeq N
\end{eqnarray}
where $g(r) = (\xi(r)/a)^{5/3}$ is the number of segments in a blob at $r$.

Since $D<<R^{(def)}$, the blob size at the entrance ($r = D$) is readily approximated as 
\begin{eqnarray}
\xi (D) \simeq \frac{k_BT a}{qe E_{in}D}
\end{eqnarray}
We expect the smooth injection when $\xi(D) \le D$.
Therefore, we find that the injection field $E^{(inj)}$ is sparsely larger than the deformation field, but no dramatic difference is expected; $E^{(inj)} \simeq E^{(def)}$.
Thus, the injection field is almost the same as that in a slit geometry (eq. (\ref{E_cri})).
It means that once the stuck chain is deformed at the entrance, there is almost no free energy barrier left, and the smooth injection takes place (the range for the regime III is very narrow), therefore, the use of the inside approach (Sec. \ref{inside_approach}) is justified in this case.
The field $E^{(per)}$, below which the process is described by the weakly biased entropic trapping, is the same as that in the slit geometry (eq. (\ref{E_ub})).

The case with local friction picture can be worked out similarly.
For a chain obeying local friction (Rouse) dynamics, we find that
\begin{eqnarray}
E_{(R)}^{(def)} \simeq \frac{k_BT a}{qe D^2} N^{-\frac{2}{5}}
\end{eqnarray}
Close to the entrance ($r<<R^{(def)}$), the blob size is approximated as
\begin{eqnarray}
\xi_{(R)}(r) \simeq a \left( \frac{k_BT}{qeE_{in}D}\right)^{\frac{3}{5}}\left(\frac{r}{D}\right)^{\frac{3}{5}}
\end{eqnarray}
From this, we find that the injection field is given by the same scaling form as that in a slit geometry (eq. (\ref{E_inj_R}))
The field $E_{(R)}^{(per)}$, below which the process is described by the weakly biased entropic trapping, is the same as that in the slit geometry (eq. (\ref{E_R^def}); note that $E_{(R)}^{(per)} \simeq E_{(R)}^{(def)}$ for a slit geometry).

In contrast to the slit geometry, the electrophoretic mobility does not depend on the chain length in the driven biased activation regime (regime II and III) irrespective of the presence or the absence of the flow effect.
The length dependence is expected to show up in the regime I only, where the longer chains migrate naturally much more slowly.
In gel electrophoresis, this entropic trapping process is observed, when the pore size is relatively large and electric field is rather weak.
Here, one comment might be useful.
In gel electrophoresis, the entropic trapping is replaced by the reptation process at the electric field higher than a certain threshold.
Experiments show that this threshold is lower for lower gel concentrations, but seems not depend on the chain length of the migrating polymer, which accords with the feature in our injection field\cite{gel_ep_ET2}.
Of course, this comparison should be regarded as very qualitative, but the trend is apparently the same.

\section{Connection to the Flow-injection Problem}
\label{connection_flow_injection}
Polymers can be driven into a narrow slit or a capillary by applying a flow of the solvent\cite{Sakaue_confinement,deGennes_review,Sakaue_EPL,Daudi}.
Here we breifly note the connection between electric and flow induced injection problems.
The control parameter in the latter case is the current of the solvent flow $J$ ($m^3/sec$) through the pore.
Near the entrance of the pore, the flow is convergent, and the flow field is calculated as in the same way as the electric filed in Sec. \ref{SingleChainRheology} (slit) and in Appendix \ref{case_cylinder} (capillary).
The characteristic currents can be derived using the hydrodynaic-electric equivalence.

In the slit geometry, we measure $J$ as the solvent current through the slit per unit length.
From eq. (\ref{E_ub}), (\ref{E_threshold}) and (\ref{E_cri}), we find respectively,
\begin{eqnarray}
J^{(per)} \simeq \frac{k_BT}{\eta a} \left(\frac{D}{a}\right)^{\frac{2}{3}} N^{-1}
\end{eqnarray}
\begin{eqnarray}
J^{(def)} \simeq \frac{k_BT}{\eta a}N^{-\frac{3}{5}}
\end{eqnarray}
\begin{eqnarray}
J^{(inj)} \simeq \frac{k_BT}{\eta D}
\end{eqnarray}

In the capillary geometry, we measure $J$ as the overall current through cross sectional area of the pore.
From eq. (\ref{E_ub}) and (\ref{E_threshold_cylinder}), we find
\begin{eqnarray}
J^{(per)} \simeq \frac{k_BT}{\eta }\left(\frac{D}{a}\right)^{\frac{5}{3}}N^{-1}
\end{eqnarray}
\begin{eqnarray}
J^{(def)} \simeq J^{(inj)} \simeq \frac{k_BT}{\eta}
\label{J_inj}
\end{eqnarray}
The critical injection current in eq. (\ref{J_inj}) is consistent with the result obtained by different approaches\cite{deGennes_review,Daudi}.

\end{document}